\documentclass[a4paper,11pt]{article}
\usepackage{pos, ulem,comment}
\pdfoutput=1

\makeatletter
     {\bgroup\raggedright\small\section*{\refname
        \@mkboth{\MakeUppercase\refname}{\MakeUppercase\refname}}%
      \list{\name{bib\@arabic\c@enumiv}
            \@biblabel{\@arabic\c@enumiv}}%
           {\settowidth\labelwidth{\@biblabel{#1}}%
            \setlength\itemsep{0pt}
            \setlength\parskip{0pt}
            \setlength\parsep{0pt}
            \setlength\partopsep{0pt}
            \leftmargin\labelwidth
            \advance\leftmargin\labelsep
            \@openbib@code
            \usecounter{enumiv}%
            \let\p@enumiv\@empty
            \renewcommand\theenumiv{\@arabic\c@enumiv}}%
      \sloppy\clubpenalty4000\widowpenalty4000%
      \sfcode`\.\@m}
     {\def\@noitemerr
       {\@latex@warning{Empty `thebibliography' environment}}%
      \endlist\egroup}
\makeatother

\definecolor{orange}{rgb}{1.0, 0.5, 0}
\definecolor{darkgreen}{rgb}{0, 0.6, 0.6}

\def\gGF{g^2_{\text{GF}}}

\def\svev#1{\left\langle #1\right\rangle}
\def\be{\begin{equation}}
\def\ee{\end{equation}}
\def\bea{\begin{eqnarray}}
\def\eea{\end{eqnarray}}
\title{Determination of the continuous $\beta$ function of SU(3) Yang-Mills theory}

\author*[a]{Curtis T.~Peterson}
\author[a]{Anna Hasenfratz}
\author[a]{Jake van Sickle}
\author[b]{Oliver Witzel}

\affiliation[a]{Department of Physics, University of Colorado, Boulder, CO 80309, United States}

\affiliation[b]{Center for Particle Physics Siegen, Theoretische Physik 1, Naturwissenschaftlich-Technische Fakultät,
Universität Siegen, 57068 Siegen, Germany}

\emailAdd{curtis.peterson@colorado.edu}

\abstract{ In infinite volume the gradient flow transformation can be interpreted as a continuous real-space Wilsonian renormalization group (RG) transformation. This approach allows one to determine the continuous RG $\beta$ function, an alternative to the finite-volume step-scaling function. Unlike step-scaling, where the lattice must provide the only scale,
the continuous $\beta$ function can be used even in the confining regime where dimensional transmutation generates a physical scale $\Lambda_{\mathrm{QCD}}$. We investigate a pure gauge
SU(3) Yang-Mills theory both in the deconfined and  the confined phases and determine the continuous $\beta$ function in both. Our investigation
is based on simulations done with  the tree-level Symanzik gauge action on lattice volumes up to $32^4$ using both Wilson and Zeuthen gradient flow (GF)  measurements. Our continuum GF $\beta$ function exhibits
considerably slower running than the universal 2-loop perturbative prediction, and at strong couplings it runs even slower than the 1-loop prediction.}

\FullConference{%
 The 38th International Symposium on Lattice Field Theory, LATTICE2021
  26th-30th July, 2021
  Zoom/Gather@Massachusetts Institute of Technology
}

\begin{document}
\maketitle

\begin{centering}
\begin{figure}[t]
\centering
\includegraphics[width=\textwidth]{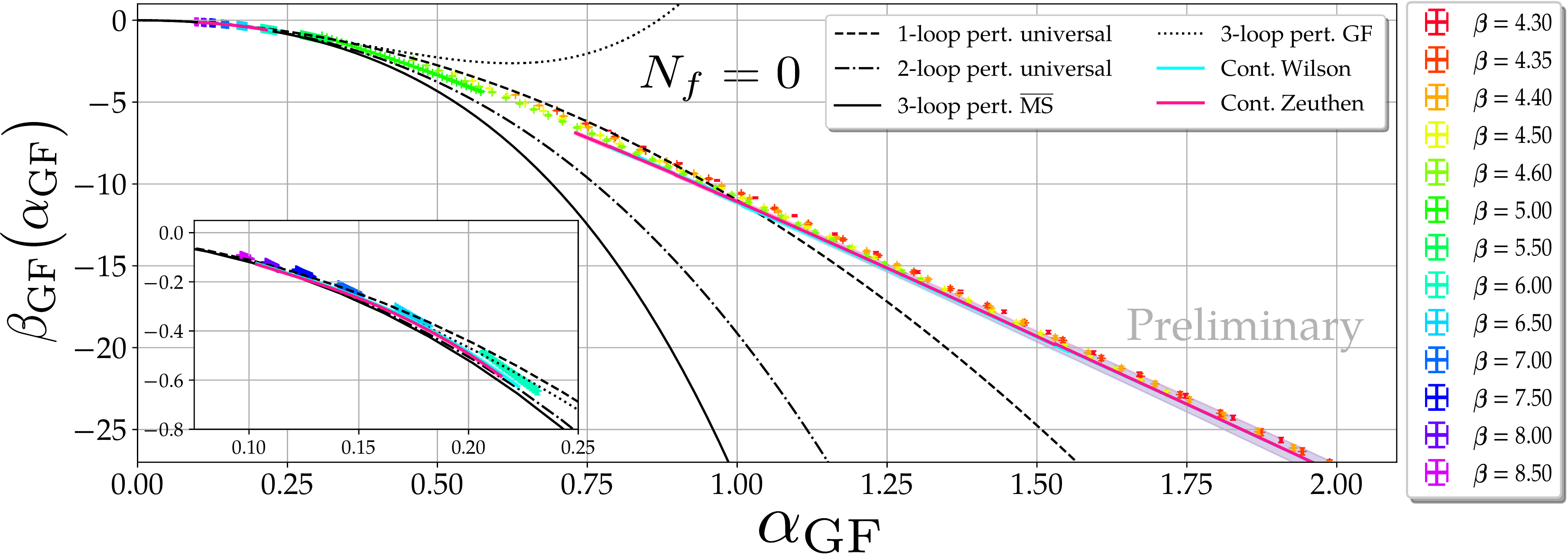}
\caption{Comparison of the continuous $\beta$ function in the continuum limit (magenta and cyan bands for Zeuthen flow and Wilson flow, respectively)
against the 1- and 2-loop universal $\beta$ functions and the 3-loop $\overline{\textrm{MS}}$ and gradient flow $\beta$ functions (various black curves).
The rainbow-colored data points correspond to ``raw data'' for the $\beta$ function on  $32^4$ volumes obtained  for several bare gauge couplings  using Zeuthen flow and the Symanzik operator (ZS) to determine $E(t)$.
The insert on the lower left shows the details in the deconfined, small $\alpha_{\mathrm{GF}}$ regime.}
\label{fig:final_result}
\end{figure}
\end{centering}

\section{Introduction}

The scale-dependent properties of a 4-dimensional Yang-Mills (YM) theory (with or without fermions) can be extracted from its renormalization group (RG) $\beta$ function, defined as the logarithmic derivative of the renormalized coupling with respect to an energy scale.
The gradient flow (GF) transformation \cite{Narayanan:2006rf,Luscher:2009eq, Luscher:2010iy} combined with Monte Carlo renormalization group (MCRG) \cite{Swendsen:1979gn}
principles defines a complete RG scheme if the energy scale is set by the GF flow time as $\mu \propto 1/\sqrt{8t}$ for asymptotically large $t$. For a detailed discussion see Ref. \cite{Carosso:2018bmz}. The GF coupling defined in terms of the flowed energy density $E(t)$
\cite{Luscher:2010iy,Fodor:2012td}

\begin{equation}
\label{GF}
\gGF(t) \equiv \frac{128\pi^2}{3(N_c^2-1)}\
\svev{t^2 E(t)}
\end{equation}
 has both zero canonical and  anomalous dimensions and therefore serves as a valid definition of a renormalized running coupling. The constants in Eq.~(\ref{GF}) are chosen to match the $\overline{\mathrm{MS}}$ coupling at tree level \cite{Luscher:2010iy} and $N_c$ denotes the number of colors. The corresponding GF $\beta$ function in infinite volume is
\begin{equation}\label{eqn:cont_beta_function}
\beta\big(g_{\mathrm{GF}}^2\big)\equiv \mu^2\frac{\mathrm{d}g^2_{\mathrm{GF}}}{\mathrm{d}\mu^2}=-t\frac{\mathrm{d}g_{\mathrm{GF}}^2}{\mathrm{d}t}.
\end{equation}

If the YM theory contains fermions, the fermion mass  would be a relevant parameter. Unless it is set to zero, its running affects $g^2_{\mathrm{GF}}$. Equation (\ref{eqn:cont_beta_function}) therefore defines the standard RG $\beta$ function only in  the chiral limit.  Attempts to determine the RG $\beta$ function using lattice simulations are usually performed  with $a m_f=0$   in small volumes in the deconfined regime, where simulations with massless fermions are feasible.  However, in the confined regime finite mass simulations might be necessary. In that case the chiral $a m_f \to 0$ limit has to be taken first \cite{Fodor:2017die}.

 Similarly, the volume is a relevant parameter. If it is kept finite in physical units,
the Callan-Symanzik RG equation contains a term proportional to $L \mathrm{d} g^2 / \mathrm{d}L$. The volume dependence of $g^2_{\mathrm{GF}}$ is non-trivial, and absorbing $L \mathrm{d} g^2 / \mathrm{d}L$ would modify the properties of the
$\beta$ function. The finite volume correction is present even if the flow time is tied to the volume, i.e.~choosing $\sqrt{8t} = c L$. In step-scaling studies this issue is avoided by scaling both the flow time and the volume at the same time. For scale change $s$ the GF step-scaling $\beta$ function is \cite{Fodor:2012td}
\begin{equation}\label{cbf}
\beta_{c,s}(u) = \frac{g^2_{\mathrm{GF}}(s^2t, s L; g_0^2) - g^2_{\mathrm{GF}}(t, L; g_0^2) }{\text{log}(s^2)}\Bigg|_{u=g^2_{\mathrm{GF}}(t,L;g_0^2)} .
\end{equation}

 To avoid introducing an unknown volume-dependent function when using Eq.~(\ref{eqn:cont_beta_function}),  the first step in obtaining the  continuous $\beta$ function  is to perform
an extrapolation to the $L/a\rightarrow \infty$ limit. We  perform the infinite volume
extrapolation at fixed lattice flow time $t/a^2$ and fixed bare coupling $\beta=6/g_0^2$. Thus  the resulting $\beta$ function corresponds to the
 $c=\sqrt{8t}/L=0$  GF renormalization scheme \cite{Fodor:2012td, Hasenfratz:2019hpg,Hasenfratz:2019puu}.\footnote{
 An alternative approach to the infinite volume extrapolation is to fix $g^2_{\mathrm{GF}}$ instead of the bare coupling $\beta$ (cf.~Sec.~\ref{numdetails}).}
The continuum physics along the
renormalized trajectory (RT) may be extracted in the second step, the ``infinite lattice flow time'' extrapolation ($a^2/t\rightarrow 0$) at fixed $g_{\mathrm{GF}}^2$.
By fixing the renormalized coupling we effectively fix the  flow time $t$.  Taking the lattice flow time $t/a^2 \to \infty$ is equivalent to the usual $a \to 0$ continuum limit and automatically tunes the bare coupling $g_0^2$ to zero.

In summary, the continuous $\beta$ function analysis follows the following steps:
\begin{enumerate}
\item Infinite volume extrapolation at fixed lattice flow time $t/a^2$ and bare coupling $\beta=6/g_0^2$.
\item Continuum limit extrapolation ($a^2/t \to 0$) at fixed $g^2_{\mathrm{GF}}$ .
\end{enumerate}

The step-scaling approach to calculating the RG $\beta$ functions on the lattice \cite{Luscher:1991wu, Fodor:2012td, Fodor:2012qh, Fritzsch:2013je, Fodor:2014cpa, Hasenfratz:2017qyr, Hasenfratz:2018wpq, Hasenfratz:2019dpr} requires that the lattice size $L$ provides the only energy scale in the system.
 In contrast the continuous $\beta$ function \cite{Fodor:2017die,Hasenfratz:2019puu, Hasenfratz:2019hpg}  requires an infinite volume extrapolation but permits the existence of a physical energy scale that might be generated by the underlying dynamics of the system.
 An example of such a scenario is the 4-dimensional pure gauge SU(3) YM  system. Here,  dimensional transmutation gives rise to an external energy scale $\Lambda_{\mathrm{QCD}}$ that characterizes
infrared behavior of the system.

In these proceedings we aim to demonstrate the properties of the continuous $\beta$ function
in the  4-dimensional pure gauge SU(3) YM system. This system has been studied extensively with step-scaling up to $g_{\mathrm{GF}}^2/4\pi\sim 1$ \cite{DallaBrida:2019wur}.   We consider volumes and bare couplings both in the deconfined (small volume, weak coupling) and in the confining (large volume, strong coupling) regimes. Figure~\ref{fig:final_result} presents our preliminary results. The rainbow colored data points show the predicted $\beta_{\mathrm{GF}}(\alpha_{\mathrm{GF}})$ as a function of the running coupling $\alpha_{\mathrm{GF}} = g^2_{\mathrm{GF}}/4\pi$ at several bare coupling values on $32^4$ lattice volumes using Zeuthen flow and the Symanzik operator (ZS)\footnote{We will define the various gradient flows, operators and other simulation details in Sect. \ref{numdetails}}. In addition the continuum limit extrapolated results both for Zeuthen (magenta line) and Wilson (cyan line) flow are shown, together with the  1- and 2-loop universal predictions and the  3-loop prediction within the GF \cite{Harlander:2016vzb} and $\overline{\mathrm{MS}}$ schemes.
A striking feature of Fig.~\ref{fig:final_result} is the smooth curve mapped out by the rainbow colored  data. These data points represent ``raw'' numerical results using Eqs.~(\ref{GF}) and (\ref{eqn:cont_beta_function}). They
 have not been altered in any way, in particular they are not extrapolated to the continuum limit. Nonetheless, these raw data points match very closely the infinite volume continuum limit extrapolated curve, indicating that the chosen action-flow-operator combination has very small cutoff effects.
This is expected in the weak coupling perturbative regime since  Zeuthen flow is $O(a^2)$ improved \cite{Ramos:2014kka, Ramos:2015baa}. Therefore the combination of Symanzik gauge action, Zeuthen flow and Symanzik operator has no tree-level $O(a^2)$ correction. However,  surprisingly, this feature persists even up to $\alpha_{\mathrm{GF}}\lesssim 2.0$, well into the confining regime.
We also observe that the $\beta$ function  follows the 3-loop GF prediction up to about  $\alpha_{\mathrm{GF}}\lesssim 0.3$, where the GF prediction peels off. The numerical result follows closely the 1-loop universal curve up to $\alpha_{\mathrm{GF}}\lesssim 1.0$,  but predicts an even slower running beyond.
In the next Section we describe our analysis in greater detail and explain how the continuum limit result shown in Fig. \ref{fig:final_result} is obtained.

\begin{centering}
\begin{figure}[tb]
\centering
\includegraphics[width=\textwidth]{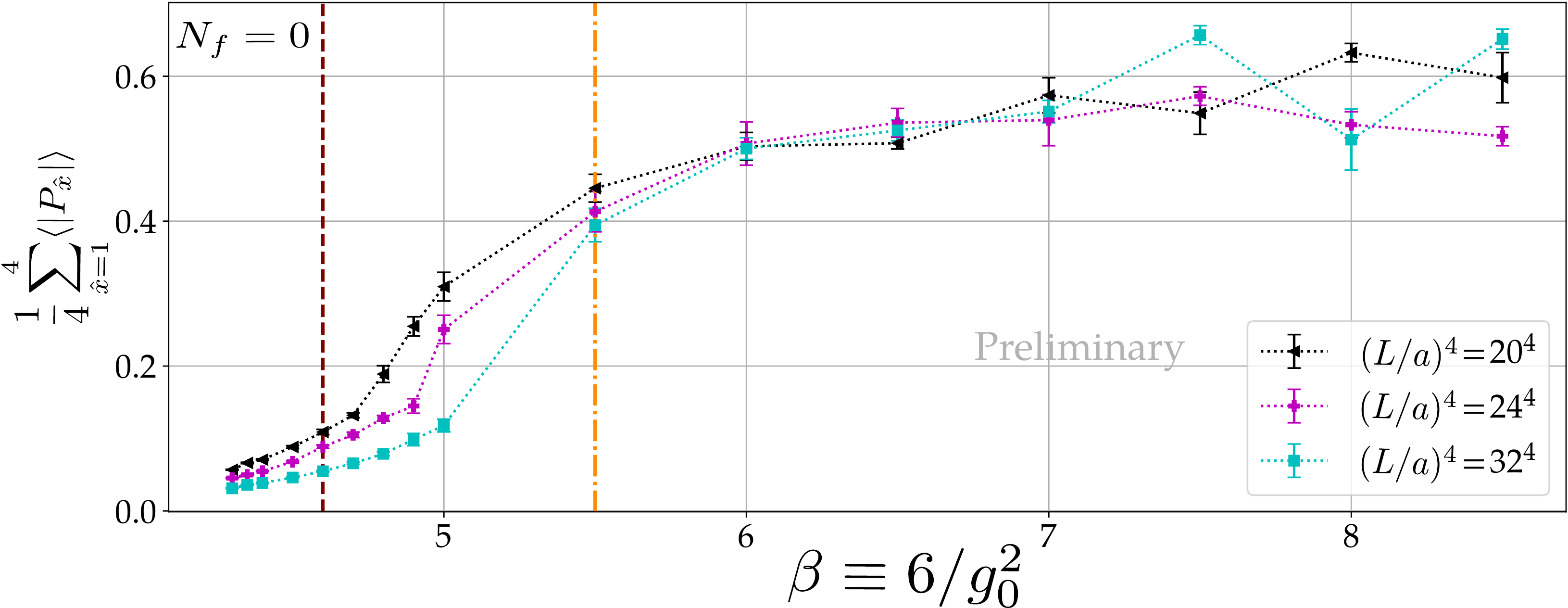}
\caption{The magnitude of the Polyakov loop as a function of the bare coupling $\beta$ for our three volumes.
 Symbols with the same color represent data
at the same volume and are connected with a dotted line to guide the eye.
Below the left dashed vertical line ($\beta\le 4.6$) all our volumes are confined while above $\beta \ge 5.5$ (right dash-dotted vertical line) all are deconfined.}
\label{fig:polyakov}
\end{figure}
\end{centering}

\section{Simulations details}\label{numdetails}

Our study is based on simulations performed using the
tree-level improved Symanzik (L\"uscher-Weisz) gauge action \cite{Luscher:1984xn, Luscher:1985zq}
and three lattice volumes of size
$(L/a)^4=20^4$, $24^4$ and $32^4$. The analysis in the confined phase is based on five bare gauge couplings ($\beta=6/g_0^2=4.3$, 4.35, 4.4, 4.5, 4.6).
In the deconfined phase, we use seven bare gauge couplings (5.5, 6.0, 6.5, 7.0, 7.5, 8.0, 8.5). The division of the simulations to confining/deconfining regimes is based on  the Polyakov loop, an order parameter in pure gauge systems. In Fig.~\ref{fig:polyakov} we show the magnitude of the Polyakov loop $P_{\hat{x}}$ (averaged
over all four directions $\hat{x}$ of the lattice) as a function of the bare coupling $\beta$ for our three volumes.  Because
simulations are performed in a finite box, we do not observe a \textit{true}\/ phase transition. However, there is a clear drop in the  magnitude of the Polyakov  loop
as a function of $\beta$ on each volume.
Scatter plots of $P_{\hat{x}}$ at each gauge configuration in the Argand plane indicate that all of our volumes with $\beta \le 4.6$ (at or below the dashed maroon vertical line)  are confined, while all
volumes with $\beta \ge 5.5$ (at or above the dash-dotted orange vertical  line) are deconfined.

We express the GF coupling as $\alpha_{\mathrm{GF}}\equiv g_{\mathrm{GF}}^2/4\pi$ and  use the definition \cite{Fodor:2012td}
\begin{equation}\label{eqn:gf_coupling}
\alpha_{\mathrm{GF}}\big(t;L, g_0^2\big) =\frac{32\pi}{3(N_c^2-1)}\frac{1}{1+\delta(t;L)}\big\langle t^2 E(t)\big\rangle ,
\end{equation}
where the factor
 $1+\delta(t;L)$ corrects for gauge zero modes that appear when imposing periodic
boundary conditions  in a finite box \cite{Fodor:2012td}.  We take the logarithmic derivative of $g_{\mathrm{GF}}^2=4\pi\alpha_{\mathrm{GF}}$  with respect to
the lattice flow time in Eq.~(\ref{eqn:cont_beta_function}) using a 5-point stencil.

Our gauge configurations are generated using \texttt{GRID} \cite{Boyle:2015tjk,GRID} and we perform gradient flow measurements
  using \texttt{QLUA} \cite{Pochinsky:2008zz,qlua}.
We consider three operators, Wilson plaquette (W), Symanzik (S) and clover (C) to calculate the energy density  $E(t)$
  using both Wilson flow (W) and Zeuthen flow (Z).  From now on we use the shorthand notation `flow-operator' to refer to a given combination, e.g.~ `ZS'  refers to Zeuthen flow and Symanzik operator.
\begin{centering}
\begin{figure}[tb]
\includegraphics[width=\textwidth]{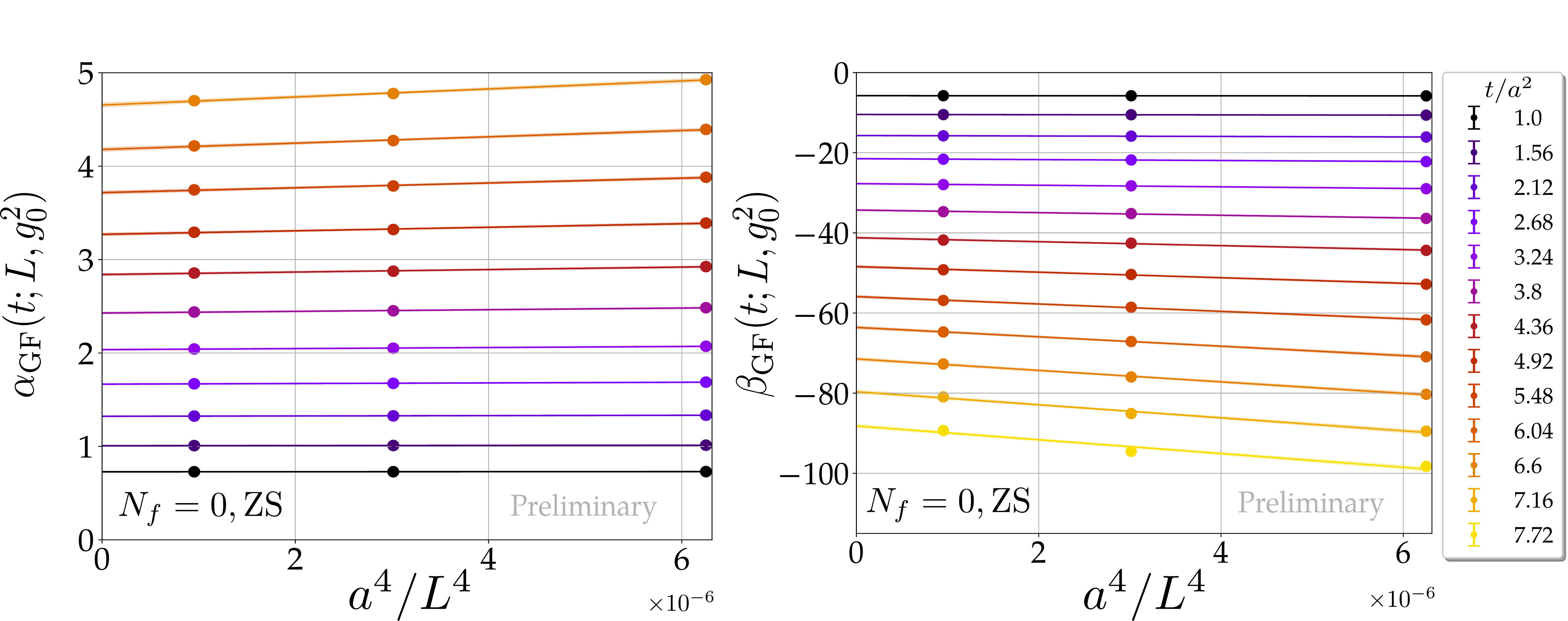}
\caption{Example of the infinite volume extrapolation for $\alpha_{\mathrm{GF}}(t;L, g_0^2)$ (left) and
$\beta_{\mathrm{GF}}(t;L, g_0^2)$ (right) at $\beta=6/g_0^2=4.3$ for the ZS combination. Flow times
range from $t/a^2=1.0$ to $t/a^2=7.72$ and the contributing lattice volumes are $(L/a)^4=20^4,$ $24^4$ and $32^4$.}\label{fig:inf_vol_extrap}
\end{figure}
\end{centering}

\subsection{Infinite volume extrapolation}

The infinite volume extrapolation  predicts $\beta_{\mathrm{GF}}$ as a function of $\alpha_{\mathrm{GF}}$ at fixed lattice flow time $t/a^2$ as $L^2/t \to \infty$. This can be achieved by interpolating the finite volume data  $\alpha_{\mathrm{GF}}(t;L, g_0^2)$ and
$\beta_{\mathrm{GF}}(t;L, g_0^2)$ in the bare coupling and taking the infinite volume limit at fixed  $\alpha_{\mathrm{GF}}$ and $t/a^2$ \cite{Hasenfratz:2019hpg, Hasenfratz:2019puu}. The interpolation is necessary in slowly running systems, but can be avoided if the running coupling evolves fast and predictions from different bare couplings show significant overlap. Here we apply the infinite volume extrapolation to \textit{both}\/ $\alpha_{\mathrm{GF}}(t;L, g_0^2)$ and
$\beta_{\mathrm{GF}}(t;L, g_0^2)$ at fixed bare coupling $g_0^2$ and lattice flow time $t/a^2$. Such a procedure is well-justified,  as long as the leading $t/L^2$ finite volume corrections are small even for the smallest volume used in the extrapolation.
Alternatively, one may first extrapolate $\alpha_{\mathrm{GF}}$ in Eq.~(\ref{eqn:gf_coupling}) to the
infinite volume limit, then plug the result into Eq.~(\ref{eqn:cont_beta_function}). We find that
both approaches agree within error over a reasonable range of $t/a^2$. However, the former method tends
to be  numerically more stable at larger  lattice flow times $t/a^2$ that are well below
the threshold for non-negligible $t/L^2$ effects.

Motivated by the finite-volume scaling behavior of the $\beta$ function
in the deconfined phase \cite{Hasenfratz:2019hpg, Hasenfratz:2019puu}
we choose our extrapolating function to be linear in $a^4/L^4$. We have also explored a number of
alternative fitting functions, including  polynomial dependencies on $a/L$,
exponential forms and ``resummed'' functional forms with a generic $k_1/(1+k_2 (a/L)^{\alpha})$ ($\alpha=2$ or 4) structure.
In all cases,  fits to $a_1 + a_2 (a/L)^4$ and $b_1/(1+b_2 (a/L)^{4})$ exhibit good $p$-values over a
wide range of lattice  flow times. The final results for both functional forms agree within errors and in these proceedings we
present results only for linear fits to $(a/L)^4$. Investigations of other possible fitting functions are still underway.
\begin{centering}
\begin{figure}[tb]
\includegraphics[width=\textwidth]{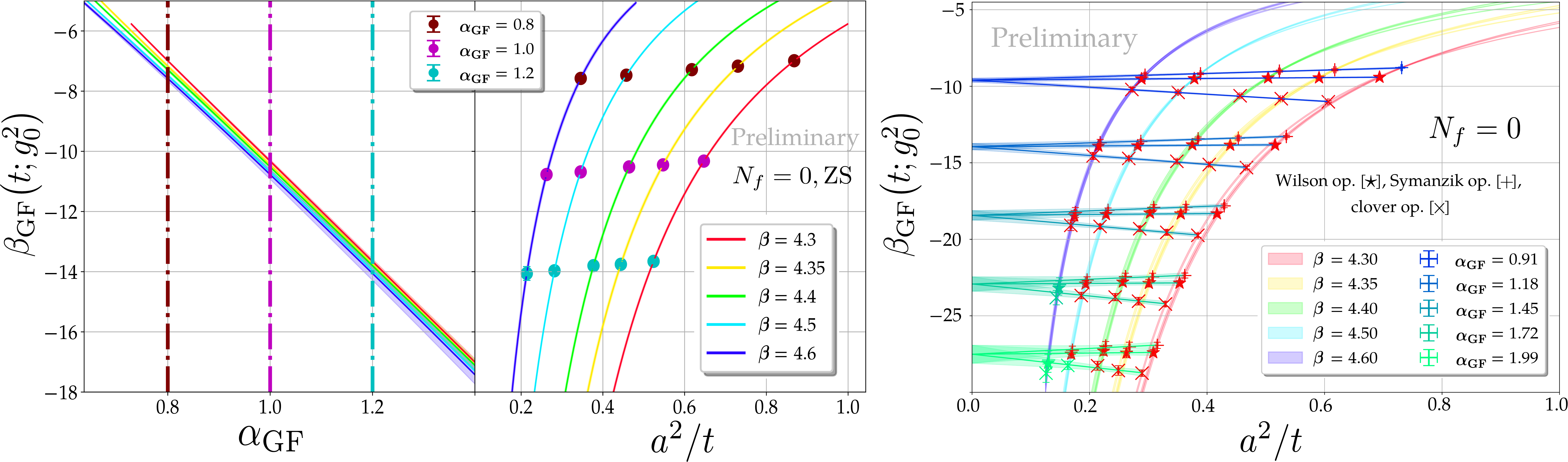}
\caption{Example of the continuum extrapolation for Zeuthen flow.  On the left side of the left panel we show the infinite volume extrapolated $\beta_{\mathrm{GF}}(t;g_0^2)$ as a function of  $\alpha_{\mathrm{GF}}$ with the Symanzik operator. At sample values of $\alpha_{\mathrm{GF}}$ (0.8 (maroon), 1.0 (magenta) and 1.2 (cyan)) we identify ($\beta_{\mathrm{GF}},t/a^2$) pairs at available bare couplings $\beta=6/g^2_0$. On the right side of the left panel  we show these pairs.
On the right panel we include the clover and plaquette operators as well and show  the results of a simultaneous linear fit to all three operators. The straight lines (blue to green) correspond to the extrapolation (with statistical errors only) at different fixed $\alpha_{\mathrm{GF}}$.
Red symbols indicate data points that were used to obtain each fit over the lattice flow time range $t/a^2\in [1.42, 6.0]$.}
\label{fig:cont_extrap}
\end{figure}
\end{centering}

Figure \ref{fig:inf_vol_extrap} shows an example of the infinite volume extrapolation at
$\beta=4.3$ using the ZS flow-operator combination. The small slopes for both $\alpha_{\mathrm{GF}}$ and $\beta_{\mathrm{GF}}$ over the entire range of
lattice flow times considered confirms  that  finite-volume effects are small in
the confined phase. Small finite-volume effects allow the use of larger lattice flow times $t/a^2$. In turn, simulations in the confined phase cover more of the
renormalized trajectory.  Finite volume effects in the deconfined phase, where the lattice size provides the only infrared cutoff, are expected to be larger. We observe that finite volume effects become significant at much smaller $t/L^2$ values in the deconfined regime
than in the confining regime. The  larger finite-volume corrections in the deconfined regime constrain the flow time values that can be used in the
 continuum extrapolation.

\subsection{Continuum extrapolation}\label{sect:contlim}

 We  determine the $\beta$ function along the renormalized trajectory  by removing the lattice cutoff (taking the
continuum limit) at a given GF coupling $\alpha_{\mathrm{GF}}$. In-between renormalization group fixed points (RGFP)
the mapping $t\mapsto \alpha_{\mathrm{GF}}(t)$ is bijective. Therefore, fixing $\alpha_{\mathrm{GF}}$
determines the dimensionful flow time $t$ uniquely.
One can  then take the continuum limit at a given $\alpha_{\mathrm{GF}}$ simply by extrapolating $\beta_{\mathrm{GF}}(t;g_0^2)$ to
$a^2/t\rightarrow 0$.

The left panel of Fig.~\ref{fig:cont_extrap}  demonstrates this procedure for Zeuthen flow and Symanzik operator. On the left side,
we fix $\alpha_{\mathrm{GF}}$ at 0.8 (maroon), 1.0 (magenta) and 1.2 (cyan). These fixed values
of the coupling will intersect the $\beta$ functions determined with different
bare gauge couplings (i.e., different lattice cutoffs), and we can ask
at what value of $t/a^2$ does each $\alpha_{\mathrm{GF}}$ intersect $\beta_{\mathrm{GF}}(t;g_0^2)$.
By collecting each of these $t/a^2$ values, we arrive at the right side of the left panel, where
we plot $\beta_{\mathrm{GF}}(t;g_0^2)$ as a function of $a^2/t$. The different colored symbols (maroon, magenta and cyan) represent
the data points at which each fixed $\alpha_{\mathrm{GF}}$ intersects $\beta_{\mathrm{GF}}(t;g_0^2)$.
We see that $\beta_{\mathrm{GF}}(t;g_0^2)$ is roughly linear in $a^2/t$, i.e.~cutoff effects scale with $a^2$.

In practice, we obtain very good results from a simultaneous linear fit to all three operators
 at fixed $\alpha_{\mathrm{GF}}$ as illustrated on the right panel of Fig.~\ref{fig:cont_extrap}. The data are highly correlated and we use an SVD cut to  tame the correlation matrix. The range of $t/a^2$ values entering the
 continuum extrapolation fit must be chosen with care. Lattice artifacts may contaminate small flow times. At large flow times residual finite volume effects can enter due to an imperfect
infinite volume extrapolation.
In the confined phase, we find that our continuum extrapolation of $\beta_{\mathrm{GF}}(t;g_0^2)$  is consistent with a
 linear dependence  over $t/a^2\in[1.42, 6.0]$. Removing the first or last bare gauge couplings that contribute to the
continuum limit does not significantly impact the central value, but,  depending on the number of data points used, could  affect the errors.

Despite each flow having different cutoff effects, different flows
should give the same continuum limit. Therefore, as a consistency check,  we compare  predictions of different gradient flows against each other to ensure they agree. This is indeed the case, as indicated by the overlap of the
cyan and magenta curves in Fig.~\ref{fig:final_result}.

\section{Conclusions and Further Prospects}

We have calculated the continuous $\beta$ function in the deconfined and confined phases of a pure gauge
Yang-Mills system using the gradient flowed gauge coupling  to determine the renormalized running coupling. The advantage of the continuous $\beta$ function over the more traditional step-scaling function is that it can be used even in the confining phase where dimensional transmutation introduces a physical energy scale.  In addition we demonstrate that
large volume simulations in a fully $\mathcal{O}(a^2)$ improved setup have the potential to provide a
good approximation of the renormalized trajectory of a 4-dimensional renormalized Yang-Mills system.

In the future, we aim to further explore alternative approaches to extrapolating to the infinite volume limit.
 Moreover, we would like to predict the continuous $\beta$ function in the transition region between deconfined and confined regimes.  We aim to explore the use of our
setup to extract the $\Lambda$-parameter and will also compare our findings to alternative approaches (see e.g.~\cite{DelDebbio:2021ryq, Francesconi:2021jcp}).
\section*{Acknowledgments}

We are very grateful to Peter Boyle, Guido Cossu, Anontin Portelli, and Azusa Yamaguchi who
develop the \texttt{GRID} software library providing the basis of this work and who assisted us in installing
and running \texttt{GRID} on different architectures and computing centers. A.H. acknowledges
support by DOE grant DE-SC0010005.  This material is based upon work
supported by the National Science Foundation Graduate Research Fellowship Program under Grant
No.~DGE 2040434. Computations for this work were carried out in part on facilities of the USQCD Collaboration, which are funded by the Office of Science of the U.S.~Department of Energy, the RMACC Summit supercomputer \cite{UCsummit}, which is supported by the National Science Foundation (awards No.~ACI-1532235 and No.~ACI-1532236), the University of Colorado Boulder, and Colorado State University. The RMACC Summit supercomputer is a joint effort of the University of Colorado Boulder and Colorado State University. We thank  BNL, Fermilab,  Jefferson Lab,  the University of Colorado Boulder, the NSF, and the U.S.~DOE for providing the facilities essential for the completion of this work.

{\small
  \bibliography{bibliograph}
  \bibliographystyle{JHEP-notitle}
}

\end{document}